%% file: scam.tex
\documentclass[sigconf,nonacm]{acmart}

\setcopyright{none}

\usepackage[table]{xcolor}
\usepackage{enumitem}
\usepackage{graphicx}
\usepackage{booktabs}
\usepackage{multirow}
\usepackage{array}
\usepackage{mathtools}
\usepackage{xurl}
\usepackage{graphicx}
\usepackage{url}

\usepackage[most]{tcolorbox}

\DeclareGraphicsExtensions{.pdf,.png,.jpg,.jpeg}
\graphicspath{{./figs/}}

\title{Read This Paper to Get \$50 Million:\textsuperscript{\textasteriskcentered} An Analysis of Mobile Messaging Scams Using Reddit Data}

\author{Allison Lu}
\email{allison.lu@ufl.edu}
\affiliation{
  \institution{University of Florida}
  \city{Gainesville}
  \state{Florida}
  \country{USA}
}

\author{Bernardo B. P. Medeiros}
\email{b.broetopontimed@ufl.edu}
\affiliation{
  \institution{University of Florida}
  \city{Gainesville}
  \state{Florida}
  \country{USA}
}

\author{Kevin R. B. Butler}
\email{butler@ufl.edu}
\affiliation{
  \institution{University of Florida}
  \city{Gainesville}
  \state{Florida}
  \country{USA}
}

\author{Patrick Traynor}
\email{traynor@ufl.edu}
\affiliation{
  \institution{University of Florida}
  \city{Gainesville}
  \state{Florida}
  \country{USA}
}

\begin{document}
\emergencystretch 3em

\begin{abstract}
\input{abstract}
\end{abstract}

\maketitle

\begingroup
\renewcommand\thefootnote{\fnsymbol{footnote}}
\footnotetext[1]{This paper does not provide any monetary reward. The title references a scam script identified in our dataset and is not an actual offer.}
\endgroup

\input{intro}

\input{background}
\input{methodology}
\input{characterization}
\input{content_analysis}

\input{pn_analysis}
\input{measuring_solutions}
\input{limitations}
\input{relatedwork}
\input{conclusion}

\bibliographystyle{ACM-Reference-Format}
\bibliography{bib}

\appendix
\input{appendix}

\end{document}

%% file: abstract.tex
\label{abstract}Mobile messaging scams--fraudulent messages delivered over SMS and other mobile applications--have become a persistent and evolving security threat, yet the attributes underlying these campaigns remain unclear. This study seeks to address this gap by examining trends in mobile messaging scams and testing the effectiveness of commercial and open-source off-the-shelf detection tools. We characterize mobile messaging scam operations, focusing on how phone numbers, URLs, and text content are used across campaigns. To achieve this objective, we collect and measure a dataset of 175,430 user-reported mobile messaging scams from Reddit between June 2020 and December 2025. While reply-based scams constitute only 50\% of our dataset, their compound annual growth rate (99.98\%) is nearly twice that of click-based scams (57.29\%). Critically, reply-based scams also show the lowest detector performance--despite identifiable similarities in text content and phone number origin within categories--indicating that current off-the-shelf tools are ineffective. These results suggest that further development of detectors is necessary to defend against this rapidly changing ecosystem. By examining a range of message attributes, this work provides new insights into mobile messaging scams, informing the design of more targeted and robust detection methods.



%% file: intro.tex
\section{Introduction}
\label{intro}

Mobile phones are by far the dominant platform for computing around the
world~\cite{Costin_2019}. Whether for entertainment and social media, or business
communications and payments, these systems allow users to interact with
the larger digital and physical worlds regardless of the traditional
infrastructure surrounding them. As such, these devices represent a
crucial mechanism for improving the lives and livelihoods of people
around the world.

Given the widespread use of this platform and the transformative nature
of the applications they can run, it is no surprise that they have
become increasingly targeted by malicious parties. While traditional
email spam has been readable on this platform for decades~\cite{Dhah_2019}, new
and sophisticated campaigns targeting mobile messaging applications
(e.g., SMS, Over-The-Top applications such as WhatsApp and Telegram,
etc) have anecdotally become common occurences for nearly every user.
While virtually all users seem to have experienced such mobile messaging
scams~\cite{Grauer_2025}, the longitudinal study of the structure, tactics, and
content of such messages has not yet been conducted.

In this paper, we perform the first such study. Specifically, we gather
over five and a half years of English language scam messages from relevant communities on
Reddit and organize them according to the Federal Trade Commission's (FTC) categorizations~\cite{FTC_2025}. In so doing, we make the following contributions:

\begin{itemize}[noitemsep,topsep=0pt,parsep=0pt,partopsep=0pt]
    \item {\bf Large-Scale Dataset of Mobile Messaging Scams:} We collect and process a dataset of 175,430 user-reported mobile messaging scams posted between June 2020 through December 2025, spanning SMS and several messaging applications, which provides the first extensive corpus of mobile messaging scams.

    \item {\bf Measurement and Characterization of Mobile Messaging Scam Operations:} We analyze mobile messaging scam content and attributes (e.g., phone numbers and URLs) using user-reported scams from Reddit, identifying patterns in scam scripts, category-specific behaviors, phone number origins, and URL infrastructure. Our analysis shows that many scams follow rigid templates, phone numbers originate from 106 countries, and all scam categories employ URL obfuscation techniques, most commonly shorteners.

    \item {\bf Evaluation of Commercial and Open-Source Detection Tools:} We assess widely deployed link- and text-based detection systems (i.e., VirusTotal, several LLMs) and show that reply-based scams are systematically underdetected, revealing areas for improvement in current off-the-shelf defenses. We find that all URLs show high levels of misclassification using antivirus tools. In addition, we also show that LLMs currently struggle to accurately classify mobile messaging scam texts, suffering from false positives (FPs) as high as 60.2\% and false negatives (FNs) as high as 18.8\%.

\end{itemize}

While recent works~\cite{Sims_2025, Agarwal_2025} have analyzed aspects of the mobile messaging scam ecosystem, existing analysis is limited in scope. In addition, the efficacy of current off-the-shelf defenses used in other forms of scam messaging (e.g., phishing), is not clear. Recent works on mobile messaging scams characterize specific scam types (e.g., `hi mum and dad' scams) by examining 711 interaction logs generated in conversations with scammers~\cite{Agarwal_2025} and perform thematic analysis on 1,525 Reddit posts and comments~\cite{Sims_2025}. 
Our large-scale measurement study represents a unique contribution that expands analysis beyond these existing efforts by two orders of magnitude. 

The remainder of this paper is organized as follows:
Section~\ref{sec:background} provides background on the forms and origins of mobile messaging scams;
Section~\ref{sec:methodology} introduces our research questions and describes the methodology used to collect and process the dataset analyzed in this measurement study;
Section~\ref{sec:characterization} presents a characterization of our mobile messaging scam dataset;
Section~\ref{sec:content_analysis} examines similarities and differences in text content across scam categories and subcategories;
Section~\ref{sec:pn_analysis} analyzes phone number origins and their associations with specific scam campaigns;
Section~\ref{sec:measuring} evaluates the performance of current off-the-shelf detection tools in identifying mobile messaging scams;
Section~\ref{sec:limitations} discusses findings and limitations;
Section~\ref{sec:relatedwork} reviews related work on scam characterization and detection;
Section~\ref{sec:conclusion} provides concluding remarks.

%% file: background.tex
\section{Background}
\label{sec:background}
Messaging scams include a wide variety of strategies, but they can be broadly understood in terms of the intention of the scammer's initial message. In this section, we provide context regarding categories of mobile messaging scams defined in previous literature and through the FTC's descriptions of mobile scams~\cite{FTC_2025}. Some scams are \textit{click-based}~\cite{Pearce_2014}, attempting to extract value immediately by directing recipients to malicious links~\cite{Falade_2023}; other scams are \textit{reply-based}~\cite{Acharya_2024}, focusing on drawing victims into conversation, establishing trust before escalating to requests for money or information~\cite{Agarwal_2025}. This distinction between reply-based and click-based scams is central to our analyses, as it reflects two fundamentally different attacker intents and reflects how scams evolve, spread, and monetize. These categories should be used as general classifications and guidelines. Categories can further intersect or diverge from each other based on user interactions

\subsection{Reply-Based Scams}
\begin{figure}
    \centering
    \includegraphics[scale=0.29]{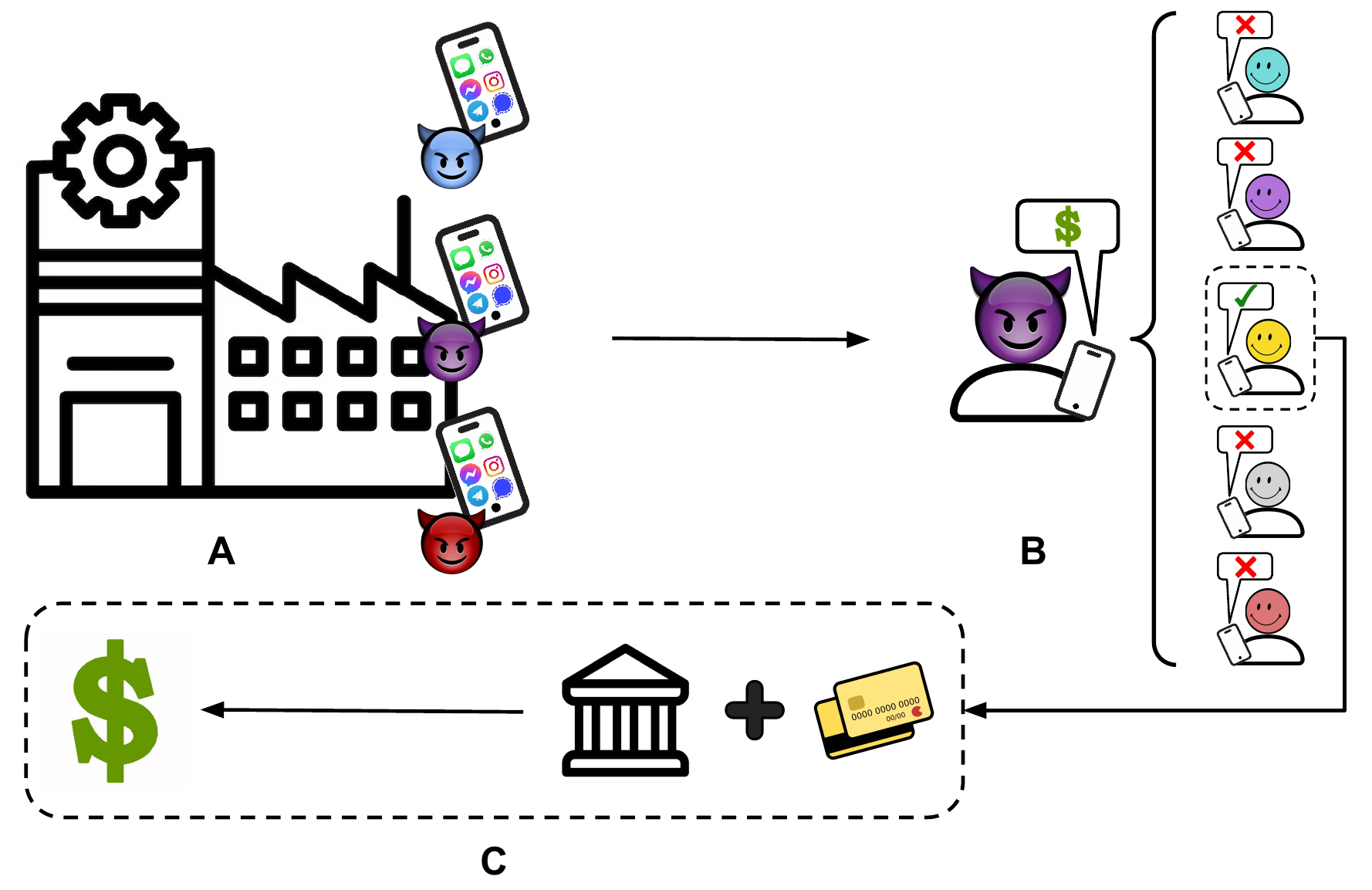}
    \caption{Numerous scam campaigns originate from scam farms (A), sending fraudulent messages at a large scale. Scammers are given accounts on various messaging platforms (e.g., iMessage, Instagram, Messenger, Signal, Telegram, WhatsApp). Scammers may use reply- or click-based tactics in their messages to recipients (B), ultimately leading to financial exploitation, where scammers obtain the victim's bank or personal information (C).}
    \label{fig:scam_pipeline}
\end{figure}

Reply-based scams use a slower and more manipulative approach. Rather than pushing links immediately, they aim to elicit a response and establish rapport with the victim, creating a false sense of trust before requesting personal or financial information (often in the form of cryptocurrency)~\cite{Han_2023,Reuters_2024}. These scams require sustained engagement to maintain conversations over time. Scammers may use a variety of scripts and social engineering techniques to manipulate their victims through phone numbers and messaging applications (as shown in Figure~\ref{fig:scam_pipeline}), often adapting in real-time based on the victim's responses. Although more resource-intensive, these scams can yield higher-value payoffs~\cite{FTC_2025,Kan_2025}. Reply-based scams have also received growing media coverage in recent years~\cite{Robles_2023}, shaping public awareness of messaging scams.

Commonly discussed reply-based scams in prior work include:
\begin{itemize}[noitemsep,topsep=0pt,parsep=0pt,partopsep=0pt]
    \item \textbf{Fake Job:} Scammers pose as recruiters or legitimate employers, often using the names of real companies or fake job postings~\cite{Ravanelle_2022}. Victims may be asked to pay “application fees,” purchase equipment up front, or provide sensitive personal information such as Social Security numbers and bank details for identity theft~\cite{Prashanth_2022}. 
    
    \item \textbf{Romance:} Attackers pose as romantic partners, manipulating victims into sending money or sensitive information~\cite{Han_2023}. Scammers often use photos stolen or generated from the web, engaging in conversations with their victims for extended periods of time~\cite{Coluccia_2020}. 

    \item \textbf{Wrong Number:} These scams begin as casual, misdirected messages that transition into long-term conversations that lead to financial or personal exploitation~\cite{Agarwal_2025b}.
\end{itemize}

\subsection{Click-Based Scams}

Click-based scams~\cite{Sadeghpour_2021} aim to exploit the user in a single message--through the use of links, domains, and hosting services--to achieve rapid monetization~\cite{Pearce_2014}. These scams can generally be considered the mobile versions of classical email phishing attacks. Click-based scams typically impersonate legitimate organizations and rely on user's trust or sense of urgency to lure them into clicking links that lead to credential theft, financial fraud, or installing malware on victim devices~\cite{Sadeghpour_2021}. 

Unlike reply-based scams, click-based scams depend heavily on scalable infrastructure rather than sustained interaction with victims. These scams require the registration of domains, hosting of malicious sites, and access to bulk messaging tools for distribution. Their effectiveness relies on rapid deployment and short-lived infrastructure to evade detection. Prior work has described several recurring click-based scam types, with the most-discussed categories listed below:

\begin{itemize}[noitemsep,topsep=0pt,parsep=0pt,partopsep=0pt]
    \item \textbf{Account Verification/Payment:} These messages impersonate banks or service providers, directing recipients to spoofed websites designed to harvest login credentials or personal information~\cite{Sadeghpour_2021}.
    \item \textbf{E-Commerce:} Scams impersonating Amazon or other online sellers, using a wide range of scripts. They may claim that an order or delivery requires confirmation of address or payment details, or that the recipient has received a gift from Amazon or another retail site~\cite{Falade_2023}. 
    \item \textbf{Gift/Prize:} Scams that inform recipients that they have won a prize or reward, requiring them to follow a link in order to claim it~\cite{Falade_2023}.
    \item \textbf{Postal:} Although Postal scams can be considered a part of the E-Commerce category, they are often given a separate distinction~\cite{Agarwal_2025b,Lee_2025}. These messages impersonate global postal services, instructing recipients to correct an address or pay fees via malicious links~\cite{Agarwal_2025b}.
    \item \textbf{Toll/DMV:} Toll/DMV scams appear as toll service messages that impersonate real transportation agencies~\cite{Munny_2025}. These scams use fraudulent links embedded within messages to coerce victims into making fake fine payments~\cite{Smith_2025}.

\end{itemize}
These eight scam types capture the general tactics of scammers, providing the analytic framework for this study. In the sections that follow, we return to these categories repeatedly—examining how they evolve over time, how they differ in their reliance on operational infrastructure, and how they reflect broader messaging scam behaviors. 

\subsection{Scam Origins}
Many large-scale fraud operations can be traced to organized scam labor farms (industrial-scale operations where workers are often coerced or trafficked into executing fraud campaigns~\cite{Robles_2023}). These operations operate within a broader scam ecosystem (depicted in Figure~\ref{fig:scam_pipeline}), spanning message writers, infrastructure suppliers, and money launderers~\cite{Robles_2023,Eyler_2024,Fishbein_2024}. Scam farms use a wide range of messaging platforms (e.g., SMS, WhatsApp, Telegram, and social media direct messages) to increase their reach~\cite{Acland_2025}.

To deliver mobile messaging scams at scale, scammers rely on widely available, low-cost infrastructure, including rented numbers, disposable SIMs, and shared hosting platforms~\cite{Krebs_2025,Oest_2018}. These services enable rapid rotation of phone numbers and URLs after detection, facilitating evasion of blocklists and filters. The ecosystem’s disposable, decentralized operations allow scams to consistently reach users’ devices despite ongoing intervention efforts by telecom providers and regulators~\cite{Fishbein_2024}.

While prior work has examined the organizational structure and infrastructure behind scam campaigns~\cite{Reaves_2016,Nahapetyan_2024}, the content and diversity of scam messages themselves remain comparatively underexplored. In particular, the composition and evolution of these messages, beyond isolated news reports, have received little systematic analysis. This work addresses this gap with a measurement-based approach that enables large-scale analysis of scam message content and delivery trends.



%% file: methodology.tex
\section{Methodology}
\label{sec:methodology}
To understand the content and operational behaviors of mobile messaging scams facing users, we conduct a large-scale characterization based on images of scams uploaded to Reddit from June 2020 through December 2025~\cite{reddit_data}. In this section, we introduce our research questions, why we use Reddit as a data source, outline our data acquisition techniques, and describe preprocessing methods, also shown in Figure~\ref{fig:methodology} and further discussed in Appendix~\ref{appendix:methodology}.
    \begin{figure}
    \centering
    \includegraphics[scale=0.4]{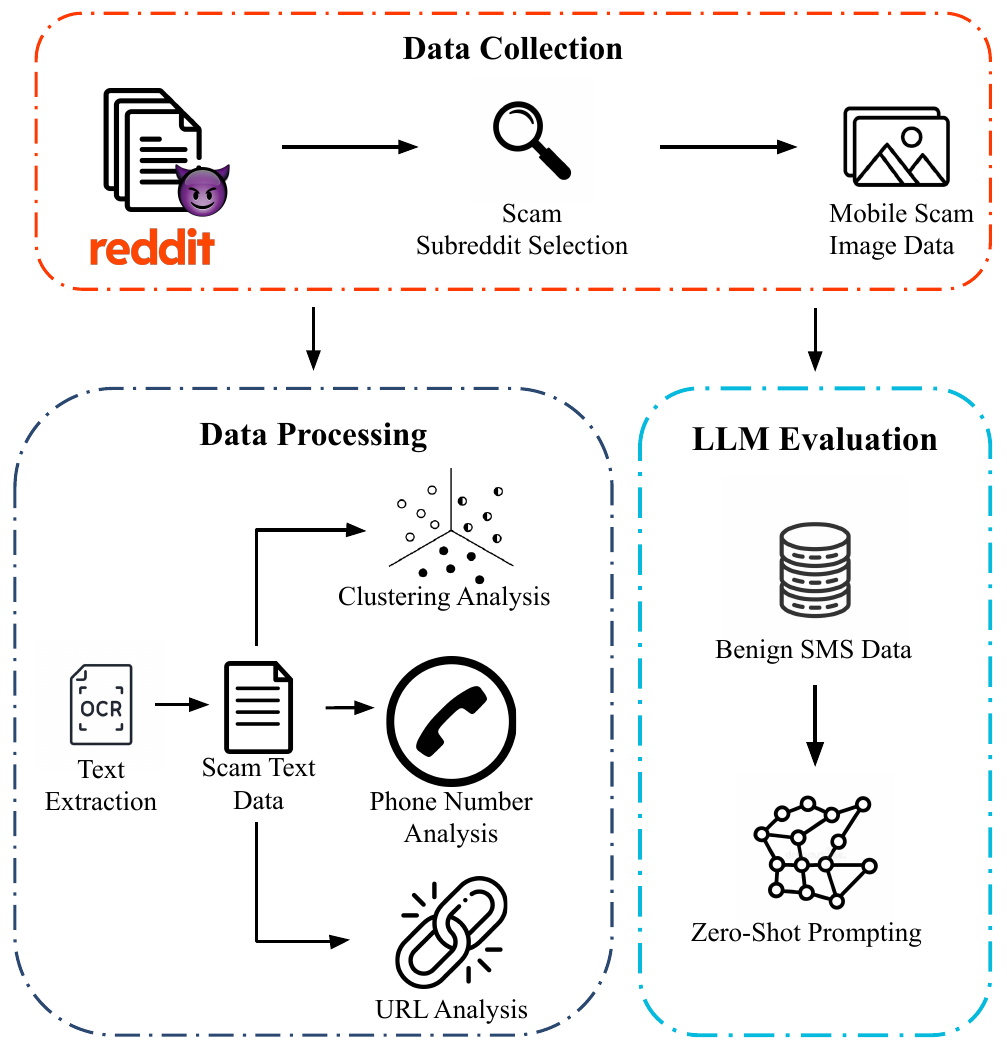}
    \caption{Diagram of our data selection and processing pipeline. We leverage pre-existing Reddit JSON data from selected subreddits. For content analysis, we convert posts to text and extract phone numbers and URLs. For LLM evaluation, we use both text and associated images from the same JSON data and apply zero-shot prompting to assess LLM accuracy in distinguishing benign from scam messages.}
    \label{fig:methodology}
    \end{figure}

\begin{figure*}
    \centering
    \includegraphics[scale=0.44]{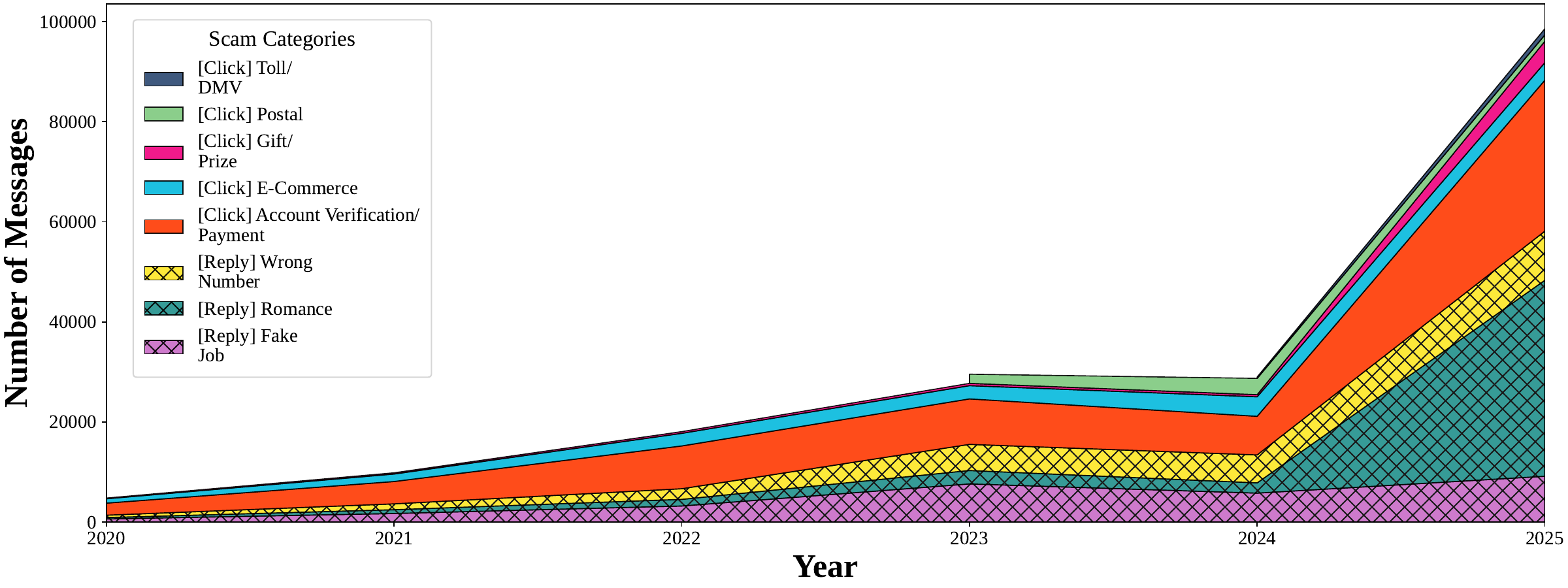}
    \caption{Year-by-year message volume by scam category, showing an increasing trend for all scams. The most common scam in the dataset is Wrong Number scams from 2020 through 2024, with a sharp increase in Romance scams in 2025.}
    \label{fig:volume}
\end{figure*}

\subsection{Research Questions}
We propose the following research questions that guide the remainder of this paper:

\begin{itemize}[noitemsep,topsep=0pt,parsep=0pt,partopsep=0pt]
    \item \textbf{RQ1:} Over the observed time period, what are the trends that characterize mobile messaging scams?
    \item \textbf{RQ2:} What content-level attributes characterize mobile messaging scams? 
    \begin{itemize}[noitemsep,topsep=0pt,parsep=0pt,partopsep=0pt]
    	\item \textbf{RQ2.1:} What attributes of URLs characterize different scam categories?
	\item \textbf{RQ2.2:} What attributes of phone numbers characterize different scam categories?
    \end{itemize}
    \item \textbf{RQ3:} Do existing off-the-shelf detection tools successfully identify mobile messaging scams at scale?
    \begin{itemize}[noitemsep,topsep=0pt,parsep=0pt,partopsep=0pt]
    	\item \textbf{RQ3.1:} How effective are antivirus products at identifying URLs used in mobile messaging scams?
	\item \textbf{RQ3.2:} How effective are state-of-the-art LLM tools at identifying mobile messaging scam text?
    \end{itemize}
\end{itemize}

\subsection{Why Reddit?}
Studying mobile messaging scams directly through telecommunications providers would provide the most comprehensive view of scam activity; however, such data is largely inaccessible due to privacy protections, encryption, and regulatory constraints~\cite{Agarwal_2025b}. Many messaging platforms use end-to-end encryption, and telecommunications providers do not publicly release message content~\cite{Chappell_2024}, making it difficult to observe scam messages at scale. As a result, we must rely on alternative data sources that provide visibility into scam activity and content from the user perspective, specifically using Reddit data.

Reddit serves as a reasonable proxy for this work because users frequently share screenshots, message transcripts, and descriptions of scam interactions when seeking advice or reporting suspicious activity. This self-reporting behavior provides access to real-world scam messages and interactions that would otherwise be difficult to obtain. Prior work has similarly used Reddit~\cite{Sims_2025,Oak_2025a,Sims_2024,Soliman_2019,Glenski_2019,Chandrasekharan_2018}, highlighting its utility as a large-scale observational dataset. Although this approach has limitations, which we discuss in Section~\ref{sec:limitations}, limited access to telecommunications infrastructure and private messaging platforms makes Reddit one of the few feasible sources for collecting longitudinal data on mobile messaging scams.
 
\subsection{Data Acquisition and Filtering}
Our Reddit dataset extends beyond individual carriers and countries, offering visibility into what users actually receive and experience. Accordingly, our dataset collection process uses historical data from archived subreddit submissions related to mobile messaging scams using the Arctict Shift API~\cite{reddit_data}. We use archived submissions as our baseline because independently gathering a dataset of similar scale would be prohibitively difficult (or even infeasible) due to Reddit’s platform restrictions, data availability limitations, and terms of service compliance~\cite{Reddit}. Our dataset is constrained to content from the top 20,000 subreddits and covers the period from June 2020 onward, due to archival gaps and broken links from older posts. Additionally, we do not include private messages or smaller subreddits. As a result, our analysis reflects scams that users (predominantly from the United States and Canada)~\cite{Shewale_Naik_2025} have chosen to publicly report, which may skew the results toward certain categories, discussed in more detail in Section~\ref{sec:limitations}.

We choose to analyze the following subreddits, which are included in this dataset: \textit{r/Scams, r/scambaiting, r/scambait,} and \textit{r/scammers}. These subreddits are chosen based on their availability, size, and relevance to our analyses. In particular, they represent the largest and most active communities dedicated to discussing, reporting, and documenting mobile scam content, ensuring that our dataset captures a broad and representative sample of real-world scam activity. We also considered the subreddit \textit{r/phishing}, but omitted it from our analyses due to the prevalence of emails over messaging screenshots. This dataset has a total of 560,708 URLs, of which 396,413 are reachable links with valid images. For each submission, we collect metadata (e.g., timestamp and image URL). Image URLs are automatically validated using web scraping, where unreachable links are discarded. Duplicate messages are also removed based on image filenames between subreddits.

To remove irrelevant content (e.g., selfies, email screenshots), we apply a two-step filtering process. First, we use email format detection to remove email screenshots. We identify and score the presence of common email patterns and wordings in each image, such as sender information (e.g., ``CC:", ``BCC:", email addresses) and the structure of emails (top-to-bottom replies instead of left-to-right exchanges, similar to those evident in text messaging). We combine this method with text message bubble detection, which consists of identifying text message-style chat bubbles and backgrounds~\cite{Dhah_2019}. If an image fails both of these tests, it is discarded from our dataset. 

\subsection{Data Processing}
Following the data collection process, we extract text from each image using Tesseract Optical Character Recognition (OCR)~\cite{ocr}. We extract text from both the entire image and just scammer-side messages. To extract only the scammer's dialogue, we maintain the image's text structure and spacing, removing anything on the sender's side based on the sender- and receiver-side spacing.

Due to variance in image structure (e.g., differences in text font, alignment, image quality, background color, and device layout), OCR performance can vary. To mitigate these inconsistencies, we perform multiple layers of image preprocessing, including conversion to grayscale, color contrast enhancement, image resizing for improved OCR resolution, removal of low-resolution images, and denoising to enhance text clarity (further discussed in Appendix~\ref{appendix:methodology}). To further validate the correctness of OCR and provide confidence in our dataset, two authors each manually reviewed and compared 100 images to their text in the dataset for $>85\%$ accuracy, due to variance between images, described in Section~\ref{sec:limitations}. As part of our filtering and image preprocessing pipeline, we remove screenshots corresponding to desktop layouts and retain only those consistent with mobile interfaces. Desktop screenshots are identified based on visual layout characteristics such as multi-column message panes and large screen aspect ratios that differ from typical mobile text messaging views. By filtering these instances, we establish that the remaining images reflect the presentation and interaction patterns of scams as they appear in mobile environments.

After capturing all the text from each image, we perform several text preprocessing steps, including text cleaning, spellcheck, normalization through lemmatization, stop-word removal, and lowercasing all text in the dataset. This process allows us to identify semantic and text similarity for further analysis tasks. We also remove empty rows, non-English text, and noise. This process effectively removed 2,156 rows of non-English text, resulting in 369,149 individual messages. Furthermore, we combine texts from continued conversations by mapping file names and orderings, and merging their texts together into a single row, resulting in 175,430 total message exchanges. Finally, we store the file in a structured comma-separated values (CSV) format for further analysis. Additional details of our methodology are provided in Appendix~\ref{appendix:clustering} and~\ref{appendix:methodology}. 


%% file: characterization.tex
\section{Data Characterization}
\label{sec:characterization}
To understand the scope, evolution, and operational characteristics of mobile messaging scams, this section characterizes the 175,430 mobile messaging scams in our dataset using categories from Section~\ref{sec:background}. This data spans five and a half years, with a focus on category distribution, temporal trends, phone number origins, and URL usage. We summarize the distribution of messages across classes and examine basic textual properties (e.g., distribution of phone numbers, URLs, and scam categories). This initial characterization provides a foundation for our subsequent analyses of the data.


\subsection{Temporal Distribution}
\begin{figure}[!htbp]
    \centering
    \includegraphics[scale=0.33]{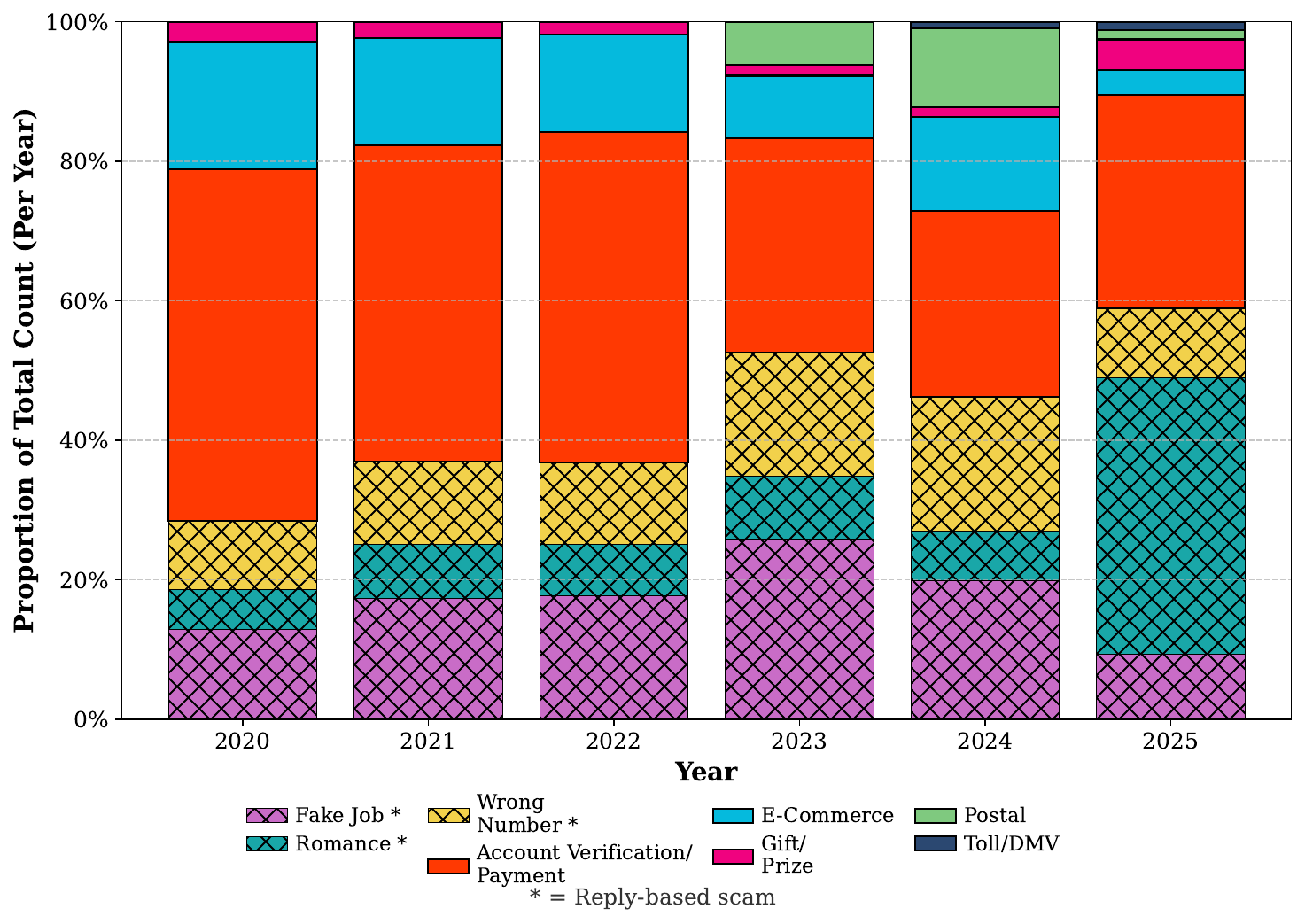}
    \caption{Reply-based scam growth (CAGR = 99.98\%) outpaces click-based scam growth (CAGR = 57.29\%). Hatched bars represent reply-based scams.}
    \label{fig:yearly_proportion}
\end{figure}

The temporal trends observed in our data from mid-2020 through the end of 2025 are shown in Figures~\ref{fig:volume} and~\ref{fig:yearly_proportion}. Early coverage (June 2020 through January 2021) is limited due to the start date of data collection. Scam volume grows steadily thereafter, peaking at 98,584 conversations in 2025. This observed increase may indicate both higher scam activity and increased user reporting.

\subsubsection{Reply-Based Scams}
Reply-based scams show the most change over time, with a category-wide cumulative annual growth rate (CAGR) of 99.98\% over the span of the dataset. Within this category, Romance scams dominate by volume (growing from 275 reports in 2020 to over 39,000 in 2024), with the highest CAGR of 167.92\%, shown in Figure~\ref{fig:volume}. Fake Job and Wrong Number scams show similar growth patterns, with Fake Job scams growing from 615 reports in 2020 to 9,145 in 2025 (CAGR of 52.23\%) and Wrong Number scams beginning with 475 to 9,816 (CAGR of 70.15\%) reports during the same period. Reply-based scams represent the fastest-growing group of scams in our dataset, characterized by substantial year-over-year changes, as shown in Figure~\ref{fig:yearly_proportion}. This trend is consistent with adversaries increasingly shifting toward conversational, rapport-building engagement rather than single messages with malicious link payloads~\cite{Sam_2025,FTC_2025}.

\subsubsection{Click-Based Scams}

Click-based scams exhibit more categorical diversity, but with lower growth rates compared to reply-based scams. The category-wide CAGR for click-based scams is 57.29\% (nearly half of the reply-based category-wide CAGR), as shown in Figure~\ref{fig:yearly_proportion}. Among persistent click-based categories, Account Payment/Verification scams show the highest sustained growth (2,411 reports in 2021 to 30,177 in 2025), with a CAGR of 61.25\%.~\footnote{The `Postal' and `Toll/DMV' categories enter the dataset late and is thus excluded from this analysis as it sees an artificially inflated CAGR due to minimal volume in early years and large spikes later in the dataset.} E-Commerce scams are the second most persistent scams, with a lower CAGR of 23.60\%, beginning with 877 and ending with 3,506 reports in 2025, shown in Figure~\ref{fig:volume}.

When separated by intention, distinct patterns emerge. Reply-based scams demonstrate faster and larger year-over-year growth, with Romance scams displaying the sharpest increase. Click-based scams, by contrast, show slower but steadier growth, with the exception of click-based Account Verification/Payment scams. In addition, the emergence of Postal and Toll/DMV scams in our dataset indicates that despite the quicker growth of reply-based scams in recent years, adversaries are still innovating new click-based mobile messaging scam scripts.

\subsection{Phone Number and URL Distribution}

Across all messages, we extract 16,745 phone numbers spanning 106 countries. Of these, 13,531 (80.79\%) belong to the North American Numbering Plan (NANP), which primarily covers the U.S., Canada, and parts of the Caribbean~\cite{fcc_nanpa_nd}. The remaining 3,218 numbers originate outside NANP regions: the Philippines appears most frequently, followed by the United Kingdom (UK) and a cluster of Southeast Asian and West African countries (in order: Nigeria, India, Indonesia, Thailand, and Malaysia). Although numbers can be spoofed~\cite{Sahin_2017}, this distribution suggests the scam ecosystem is not purely domestic; rather, many campaigns are operated or coordinated from outside the U.S./Canada, consistent with globally distributed scam operations and outsourced infrastructure~\cite{Eyler_2024,Fishbein_2024,Robles_2023,Acland_2025}.

We also extract 31,288 URLs. While well-known platforms (e.g., Google, Facebook, Apple) appear, scammers disproportionately use URL shorteners (e.g., \textit{bit.ly}, \textit{cutt.ly}, \textit{tinyurl}) and messaging endpoints (e.g., \textit{t.me} for Telegram, \textit{wa.me} for WhatsApp) to obfuscate final destinations. Critically, the top 15 domains account for only 21\% of URLs, indicating scammers distribute traffic across a long tail of domains, increasing the cost of domain-based defenses~\cite{Peng_2019}.

\begin{tcolorbox}[standard jigsaw, opacityback=0,colframe=cyan]
        \textbf{RQ1 Takeaway:} \textit{Reply-based scams, specifically Romance scams, show the most growth, increasing at nearly twice the rate of click-based scams.}
\end{tcolorbox}

%% file: content_analysis.tex
\section{Content Analysis}
\label{sec:content_analysis}
In this section, we show that while the scam landscape spans many topics and strategies, certain categories exhibit strong internal consistency, reflecting shared templates and coordinated operations. Although these subcategories can be aggregated for high-level statistics, distinguishing them reveals meaningful differences in how scams are executed and scaled. Using the FTC’s scam categorizations~\cite{FTC_2025}, we identify subcategories that rely on repeatedly reused scripts across broader scam categories. This distinction highlights which variants reflect a shared operational class and which merit separate analysis as case studies.

\subsection{Text Clustering and Categorization}
To identify patterns in mobile messaging scam content, we cluster message embeddings generated with SentenceTransformer models~\cite{sbert}, following the methodology in Section~\ref{sec:methodology}. We use k-means for its simplicity and efficiency~\cite{Ahmed_2023}. Each cluster is manually labeled using top-ranked TF–IDF keywords, forming the basis of our scam categories and subsequent structural and behavioral analyses. To validate cluster coherence and reduce over-fragmentation, two authors independently reviewed 50 messages per cluster, merging clusters with substantively overlapping content. This process ensures that resulting categories reflect meaningful, operationally distinct scam types rather than artifacts of the clustering algorithm.

    \begin{figure}[ht]
    \centering
    \includegraphics[width=0.45\textwidth]{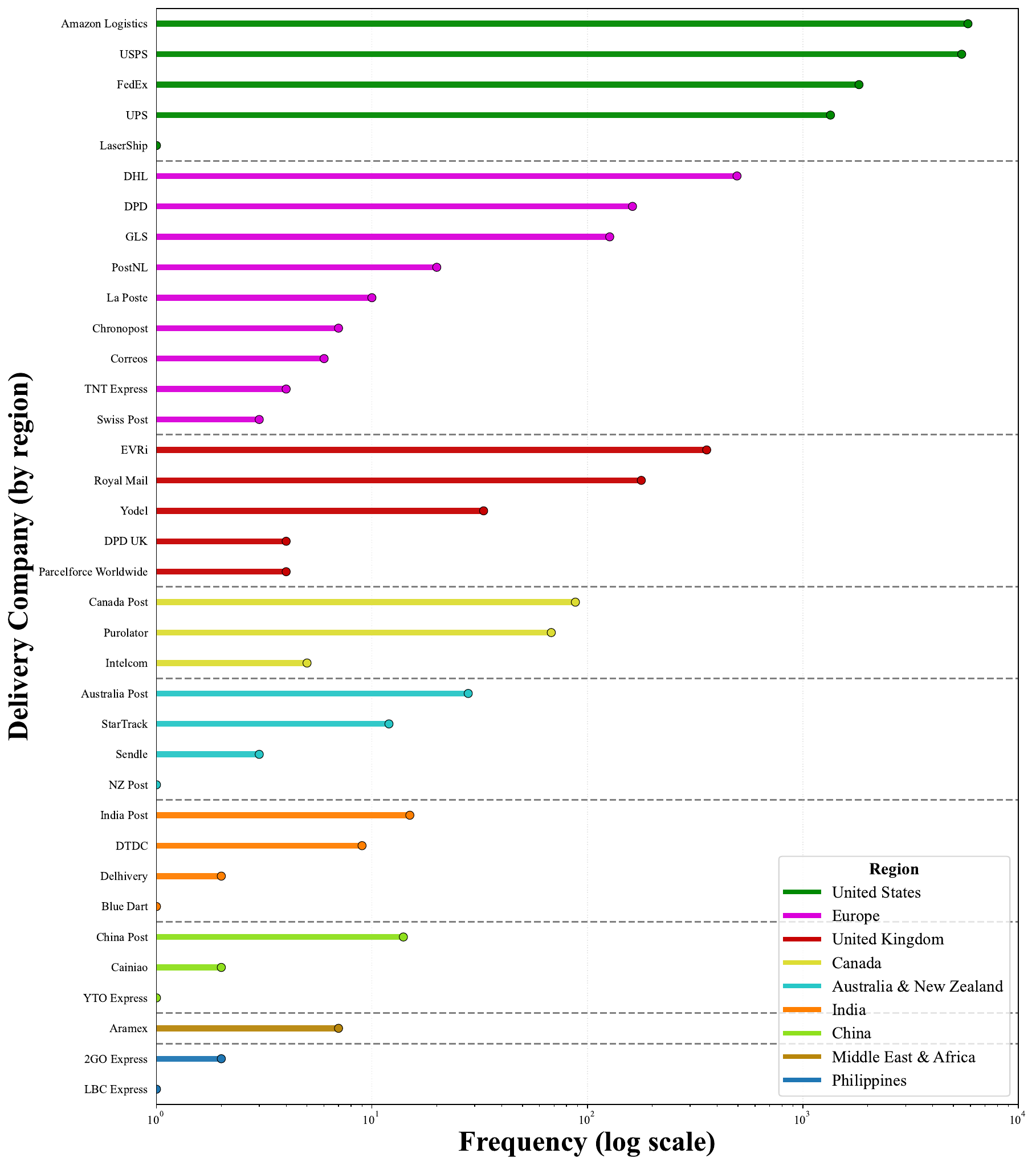}  
    \caption{Numerous global shipping/delivery companies are mentioned in our dataset, with U.S., U.K., and European companies being most prevalent.}
    \label{fig:postal_mentions}
    \end{figure}

To measure intra-category similarity, we compute mean pairwise cosine similarity, which quantifies semantic similarity as the cosine of the angle between message vectors. For each cluster, we calculate cosine similarity across all message pairs and use the mean as a representative similarity score. We perform this analysis using both SentenceTransformer embeddings and TF–IDF representations~\cite{bertopic}.

We further apply HDBSCAN~\cite{Mcinnes_2017} within each cluster to identify prominent subcategories, leveraging its robustness to noise and ability to capture irregular cluster structure. Subcluster similarity is evaluated using mean pairwise cosine similarity and the Wasserstein distance, computed by splitting each subcluster into two equal distributions and measuring the distance between them (further details in Appendix~\ref{appendix:clustering})~\cite{cosine}.

\subsection{Content Analysis Results}
We observe subcategories with highly-templated mobile messaging scam scripts in both click- and reply-based scams from Section~\ref{sec:background}. These patterns suggest opportunities for automated detection, particularly for scams that rely on rigid scripts. We describe several templated scam subcategories below, with additional details and counts found in Appendix~\ref{appendix:tables}, Table~\ref{tab:reuse}:

\begin{itemize}[noitemsep,topsep=0pt,parsep=0pt,partopsep=0pt]
    \item \textbf{Bill Scam:} This minimal one-line ``bill paid'' text containing a link is commonly found in the Account Payment/Verification category of scams in our dataset.

    \item \textbf{FasTrak Lane Scam:} The FasTrak (named after California's toll payment system) Lane scam is found in the Toll/DMV category and impersonates FasTrak, insisting that the receiver pay their fees. In our dataset, we found that these messages always use various date placeholders from December 2024 and 2025, associated with the holiday season~\cite{Rivera_2023}. 
    

    \item \textbf{Illuminati Scam:} We observe a set of ``Illuminati'' scams in the Gift/Prize category. This scam uses ``copy and pasted'' texts to victims, requiring them to pay an advance fee and to provide personally identifiable information (PII) through WhatsApp~\cite{bitdefender_nd}. We find that this scam offers the vast reward of \$50,000,000 and fame to recipients for joining the group.

    \item \textbf{Muse Scam:} This scam, also included in the Prize/Gift scam category, is prevalent in social media (primarily Instagram) direct messages. These scams are typically framed as artists offering payment for permission to use a posted photo of the receiver, then hijacking the receiver's personal information and financial details~\cite{Cost_2025}.

    \item  \textbf{Part-Time Job Scam:} We observe a variant of Fake Job scams offering part-time positions for reviewing hotels~\cite{Ravanelle_2022}. We find that these messages are long and identical between different instances. This scam also contains the longest script, beginning by stating that the victim's information was found through online recruitment agencies, then offering a job opportunity and instructions to contact the scammer via WhatsApp for further details.


    \item \textbf{Postal Address Scam:} This scam first appears in our dataset in June 2023, consistent with its initial reporting in the news~\cite{Resecurity_2023}. It forms the largest subcluster and exhibits one of the highest textual similarity in our data (96\% cosine similarity), with messages uniformly mimicking official package or mail delivery notifications. These messages claim a delivery issue and prompt recipients to click a link to reschedule delivery or obtain additional details. Despite being a relatively recent campaign, Postal scams show striking textual uniformity, suggesting opportunities for automated detection. Prior work links this scam to phishing kits that enable attackers to send iMessages via compromised Apple iCloud accounts rather than traditional SMS or calls~\cite{Resecurity_2023,Nahapetyan_2024}. The scam also impersonates a range of global logistics providers (most prominently USPS, Canada Post, Royal Mail, and EVRi) as shown in Figure~\ref{fig:postal_mentions}.
    
    \item \textbf{Reddit Modmail Scam:} These scams are a part of the Prize/Gift scam category, containing the least categorical similarity, but still following similar scripts. This scam leads Reddit users to a dating site using a link, typically ending with the phrase ``this is not a scam."

    
\end{itemize}

The introduction of highly templated scam campaigns across scam categories shows that, even as tactics evolve, scammers continue to rely on repeated content, most notably in click-based scams. Our dataset reveals a range of templated scams spanning multiple categories, including an emerging Postal scam campaign that employs nearly identical scripts derived from phishing kits~\cite{Resecurity_2023}, differing primarily in the referenced courier companies (Figure~\ref{fig:postal_mentions}). This pattern suggests that new campaigns are introduced regularly with minimal structural innovation. More recently, Toll scams have grown in prevalence, prompting public warnings advising users not to click links in these messages~\cite{Cross_2025, Collier_2025}. These scams likewise rely on highly similar, and often identical, message templates.

Message reuse is not unique to just Postal or Toll/DMV scams. Across both reply- and click-based scams, we observe repeated phrasing and high textual similarity within each scam category, extending beyond the highly similar examples discussed above. In Prize/Gift and Wrong Number scams, near-duplicate messages frequently appear across different senders, differing only in minor details such as names or synonyms. These findings indicate that while new scam campaigns continue to emerge, their internal message structures change little—likely reflecting the efficiency of reusing proven scripts and the absence of strong content-based filtering. Despite these clear structural regularities, current detection systems remain inadequate at identifying mobile messaging scams, which we show in Section~\ref{sec:measuring}.

Although we observe similar and reused scripts within certain scams, the majority of scams in our dataset are still diverse; the above-discussed templated messages represent only $<$ 5\% of total messages from our dataset. While the largest reply-based category in our dataset is Romance scams, exhibited in Figure~\ref{fig:volume}, they rarely show uniform similarity. Indeed, many of the largest scam subclusters are reply-based scams with relatively low text similarity. These scams may begin similarly (discussed in Section~\ref{sec:background}), such as Wrong Number scams that open with a name and generic greeting, but as reply-based scams, the conversation evolves based on how the recipient responds, and similarity is therefore constrained.

\begin{tcolorbox}[standard jigsaw, opacityback=0,colframe=cyan]
    \textbf{RQ2.1 Takeaway:} \textit{Click-based scams exhibit consistent, uniform text patterns and make up the majority of our dataset. In contrast, reply-based scams adapt dynamically to recipient responses, which reduces message similarity across scams and makes these scams more difficult to filter.}
\end{tcolorbox}

%% file: pn_analysis.tex
\section{Phone Number Analysis}
\label{sec:pn_analysis}

  \begin{figure*}[ht]
    \centering
    \includegraphics[width=0.95\textwidth]{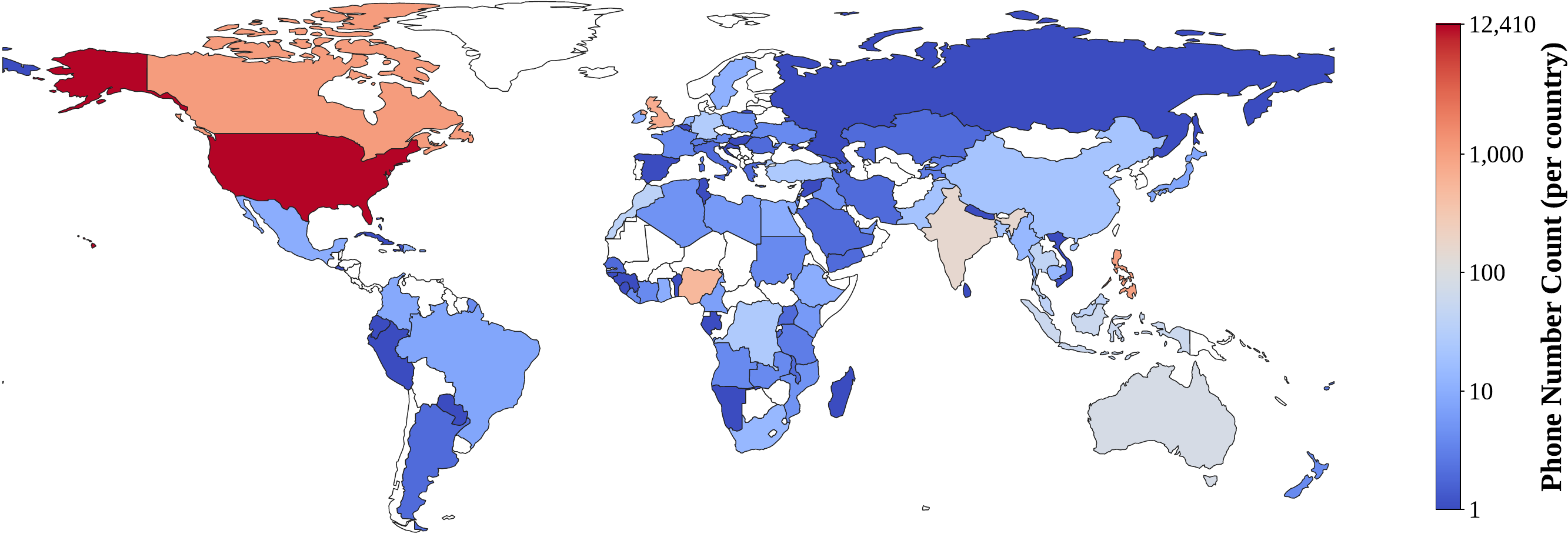}  
    \caption{Fraudulent phone number origin distribution. Most numbers are from the U.S. and Canada, but the origin distribution spans 106 countries in total.}
    \label{fig:map}
    \end{figure*}
    
In this section, we analyze country- and region-level phone number origins across scam categories and their real-world implications. Prior work has explored carrier and ownership attribution~\cite{Reaves_2016,Agarwal_2025,Costin_2019,Swetha_2025}. Given the difficulty and error-proneness of reliable ownership inference at scale~\cite{Trestle_2024}, we instead focus on aggregate origin patterns associated with scam categories (Section~\ref{sec:background}), which more robustly capture campaign-level behavior. In addition, phone number origins in scam messages \textit{do not necessarily} reflect the attacker’s physical location, as scammers frequently use number spoofing or virtual number providers. Instead, number origins often reflect strategic choices related to trust and accessibility (e.g., scammers may prefer U.S. or local numbers because they appear more trustworthy)~\cite{Verma_2024}, are easier to obtain through VoIP or SMS providers~\cite{Hart_2024}, or are less likely to be ignored than international numbers~\cite{Costin_2019}. Thus, analyzing the apparent origin of phone numbers provides insight into attacker strategies, including how scammers establish trust and select infrastructure to target specific victim populations more effectively.

\subsection{Phone Number Extraction}
To examine phone numbers and the geographic diversity of scams, we categorize each number in our dataset as either NANP or non-NANP, as described in Appendix~\ref{appendix:methodology}. We use NANP numbering as a delineator because interoperability with similar numbering schemes suggests scam content may be comparable across NANP regions~\cite{fcc_nanpa_nd}. For each cluster, we compute inter- and intra-category mean pairwise cosine similarity between NANP and non-NANP messages to assess semantic similarity across scam content originating from different regions.

\subsection{Phone Number Analysis Results}
We geographically mapped the phone numbers found from our dataset to identify country of origin and scam campaign associations. We highlight several cases of non-domestic phone numbers and their appearance in scam categories from our data. While phone numbers are diverse (shown in Figure~\ref{fig:map}), our findings suggest that many messages using NANP and non-NANP area codes share similarities in scam categories. This suggests that campaigns not only reuse similar scripts (as shown in Appendix~\ref{appendix:tables}, Table~\ref{tab:reuse}) but also share underlying attributes.

\begin{itemize}[noitemsep,topsep=0pt,parsep=0pt,partopsep=0pt]
    \item \textbf{Canada:} Canadian numbers are most strongly associated with click-based scams, particularly Toll/DMV and Postal scams, and share nearly identical content with campaigns originating from the Philippines. In Toll scams, 64\% of numbers map to Canadian area codes, consistent with reports tracing their emergence to 2024~\cite{Kendall_2025}. The remaining 36\% of numbers in the Postal scam category are also Canadian; however, these messages continue to target U.S. recipients, referencing ``USPS” and the “United States Postal Service.”
    
    \item \textbf{Nigeria:} Numbers using the Nigerian country code (+234) are primarily associated with reply-based scams, particularly Romance scams. These scams require sustained interaction and extended conversations, unlike click-based scams that do not rely on user replies. Romance scams are inherently conversational, with topics adapting to recipient responses and often culminating in a transition to other platforms, most commonly WhatsApp. We also observe Nigerian numbers in scams that begin as Romance scams that later propose investments to their victims. This combination is consistent with recent reporting on a large-scale romance and cryptocurrency operation involving Nigerian actors, in which 792 suspects were arrested~\cite{Reuters_2024}.

    \item \textbf{Philippines:} We observe a disproportionately high prevalence of phone numbers originating from the Philippines across several scam categories, particularly Toll/DMV and Postal scams. In total, 1,043 phone numbers originate from the Philippines (country codes +63 and +61) and appear in the Postal, Toll/DMV, and E-Commerce categories. Within Toll/DMV scams, FasTrak Lane scams exhibit the highest concentration, in which all identified phone numbers in this subcategory originate from the Philippines. This finding aligns with reports tracing the emergence of FasTrak Lane scams in 2024 to the Philippines~\cite{Smith_2025}. Moreover, over half (54\%) of phone numbers in this subcategory use Philippine country codes; while some numbers originate from Canada or the U.S., +63 appears most consistently~\cite{Hudspeth_2024}.
    
    \item \textbf{Thailand:} We identify 46 phone numbers originating from Thailand (country code +66), primarily associated with Wrong Number scams. These conversations exhibit low textual similarity and often begin as Wrong Number messages before transitioning into cryptocurrency investment pitches or fake trading groups. Although individual messages differ, all conversations ultimately promote fraudulent cryptocurrency or trading platforms.

\end{itemize}

The discussed countries represent the most common non-U.S. numbers from our dataset. Although the majority of numbers in our dataset are NANP numbers (shown in Figure~\ref{fig:map}), the existence of global numbers is not coincidental; the highlighted countries share the presence of English as a primary or widely spoken language with neutral tones. The use of foreign numbers to target English-speaking populations aligns with the text similarity metrics observed in our dataset: NANP and non-NANP number scams for each category consistently have approximately equal mean pairwise cosine similarity values (with only 2\%–13\% cosine similarity differences across categories), and these values remain unchanged when comparing groups. These findings suggest that NANP and non-NANP number scams are relatively similar overall, indicating that fraud networks may span phone numbers of multiple nationalities. Additionally, each of the discussed countries has been documented in existing literature~\cite{Eyler_2024,Fishbein_2024} to have experienced scam activities. This analysis further underscores the relationships between geography, language, and established fraud networks.

\begin{figure}
    \centering
    \includegraphics[scale=0.3]{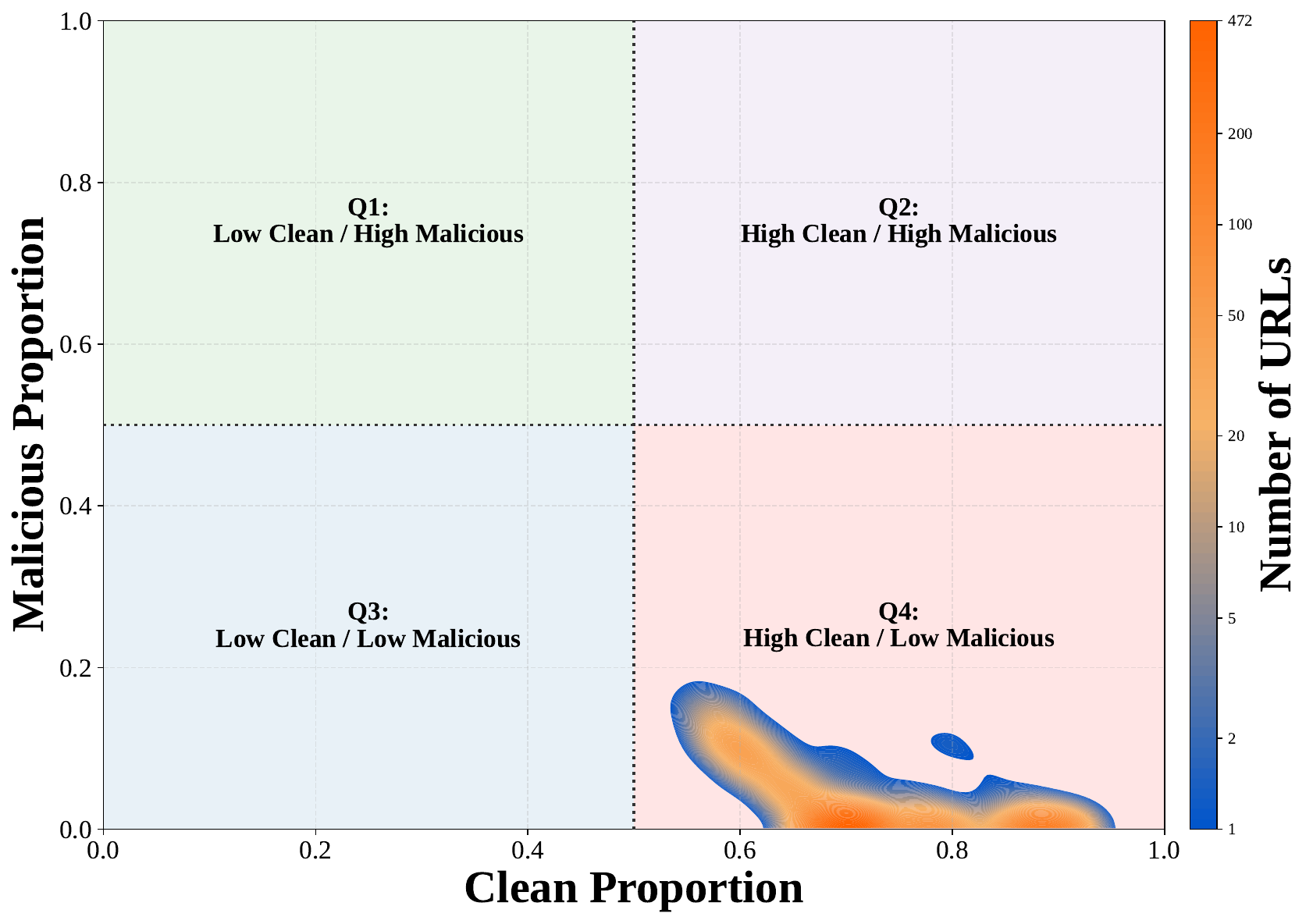}
    \caption{A majority of links in our dataset are flagged as ``clean'' (Q4) and show significant disagreement between detection engines. Ideally, link detection agreement should be in Q1 (Low Clean/High Malicious).}
    \label{fig:virustotal}
\end{figure}

\begin{tcolorbox}[standard jigsaw, opacityback=0,colframe=cyan]
    \textbf{RQ2.2 Takeaway:} \textit{Phone numbers are a key component of scam infrastructure, with Canada, Nigeria, the Philippines, and Thailand showing strong associations with specific scam categories.}
\end{tcolorbox}

%% file: measuring_solutions.tex
\section{Measuring Solutions
\label{sec:measuring}}
This section evaluates two categories of off-the-shelf/user-facing detection approaches for mobile messaging scams: (1) URL-based defenses that identify or block fraudulent links (Section~\ref{sec:url_analysis}), and (2) LLM-based classifiers that analyze message content and flag them as either fraudulent or benign messages (Section~\ref{sec:llm_detection}). Together, these analyses characterize how current default commercial and open-source tools perform against mobile messaging scams in our dataset. They also highlight detection coverage and blind spots in these tools for end-users, who often have limited literacy and understanding of how mobile messaging scams operate.
\input{url_analysis}
\input{llm_detection}

%% file: url_analysis.tex
\subsection{URL Analysis}
\label{sec:url_analysis}
\begin{figure}[ht]
\centering
\includegraphics[width=0.47\textwidth]{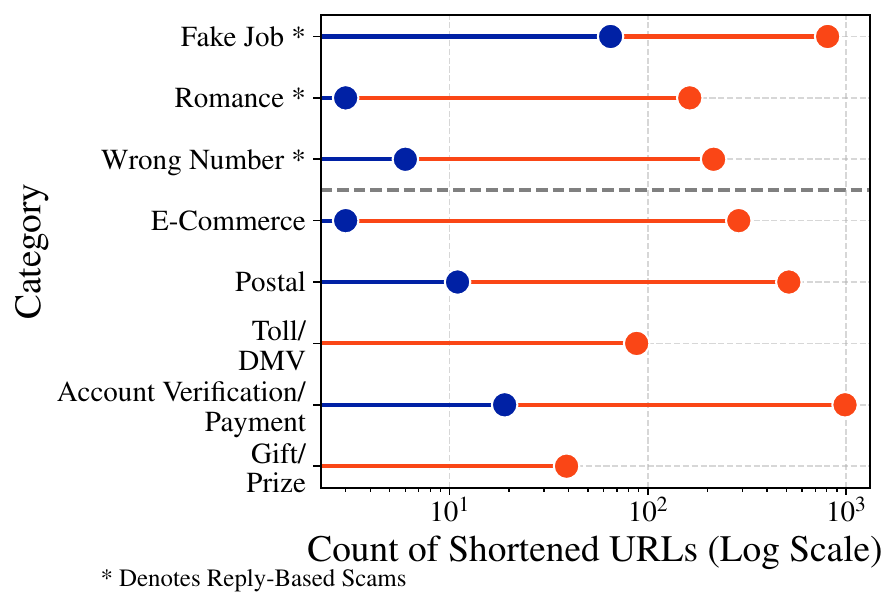}  
\caption{The click-based E-Commerce and Account Payment/Verification scams display the most active links using shortener services, while the reply-based Fake Job scam shows the most active shortened links. Blue indicates inactive links while orange represents active links.}
\label{fig:shorteners}
\end{figure}

URLs are a critical component of scam operations, serving both as communication infrastructure and gateways to identity theft~\cite{Dam_2019}. We analyze link usage and URL obfuscation in both reply- and click-based scams in this section. In reply-based scams, links typically redirect victims to encrypted messaging platforms (e.g., WhatsApp, Telegram) or obfuscated URLs (Figure~\ref{fig:shorteners}), enabling scammers to move conversations to environments where detection is more difficult. In contrast, click-based scams use URLs as the primary attack vector, directing victims to phishing pages that solicit credentials, payment details, or personal information~\cite{Dam_2019,Peng_2019}. These roles make URLs particularly valuable for understanding how scammers continue operations and move victims toward monetization.

\subsubsection{URL Extraction}

We extract 31,288 total URLs from our scam text messages dataset spanning mid-2021 through 2025 using a regular-expression-based approach described in Appendix~\ref{appendix:methodology}. Further, URLs are normalized by top-level domain, full link, and reachability. Each URL is scanned using the VirusTotal Python API, which aggregates safety assessments from over 70 antivirus engines (e.g., Avast, Bitdefender, Kaspersky), including both active and historical detections across vendors~\cite{VirusTotal}.

\subsubsection{URL Results}
We analyze URLs embedded in mobile scam messages by submitting them to VirusTotal and find substantial gaps in URL-based threat detection. Many scam-related URLs, including those that have existed for long periods of time (i.e., several years), remain undetected by a majority of antivirus engines, despite appearing in fraudulent messages~\cite{Peng_2019}. This limitation is partly due to restricted visibility into scam-specific content: antivirus resolution requests are often processed differently than links accessed via common web services (e.g., Google or social media previews), resulting in discrepancies in what detectors are able to analyze.

Comparing VirusTotal engines reveals substantial disagreement in malicious URL detection. Many URLs are flagged by some engines but not others. Notably, only a small fraction of scam-linked URLs in our dataset have ever been evaluated by VirusTotal. The majority of all links (22,344 of 31,288) have never been analyzed by any provider, and even among those that are analyzed, 64\% (5,729 of 8,944) are never flagged as malicious by any engine (Figure~\ref{fig:virustotal}). This under-reporting is even more pronounced in recent data: in 2025 alone, 90\% of links (14,651 of 16,210) have never been analyzed. Together, these results indicate that only a tiny proportion of real-world scam URLs are reported to, or detected by, widely used threat intelligence services--particularly for newer campaigns. Although some scam links may be short-lived or become benign over time, these findings highlight a lack of consensus in malicious URL detection, leaving users unable to rely on any single antivirus tool for protection. Detection is further hindered by reliance on user reports and heuristics, which often require repeated reporting or recognizable typosquatting patterns (e.g., misspellings or excessive hyphenation) before links are blocklisted~\cite{Peng_2019}.

Finally, we conduct a domain-level analysis to identify services commonly used by scammers. We find widespread reliance on URL shortening services, most notably \textit{bit.ly} and \textit{cutt.ly}, to obfuscate link destinations (Appendix~\ref{appendix:domains}). URL shorteners mask true destinations and increase the likelihood of user clicks~\cite{Page_2018}, often undermining traditional URL filtering approaches~\cite{Papez_2024}. Using a comprehensive list of shortening domains~\cite{domains}, we categorize shortened URLs by scam category and find that all scam categories employ link shorteners. Our analyses show that E-Commerce and Account Payment/Verification scams use the highest number of active shorteners, while Romance scams rely more heavily on inactive ones (Figure~\ref{fig:shorteners} and Appendix~\ref{appendix:domains}, Figure~\ref{fig:domains}). Together, these results underscore the need to prioritize proactive analysis of URLs, particularly shorteners and redirect chains, rather than relying solely on reputation-based antivirus scores that may lag behind emerging scam campaigns.

\begin{tcolorbox}[standard jigsaw, opacityback=0,colframe=cyan]
        \textbf{RQ3.1 Takeaway:} \textit{The majority of mobile messaging scam URLs are never analyzed or flagged by antivirus engines, with shortened and redirected links especially likely to evade detection.}
\end{tcolorbox}

%% file: llm_detection.tex
\subsection{LLM Analysis}
\label{sec:llm_detection} 

\begin{figure}
\centering
\includegraphics[scale=0.53]{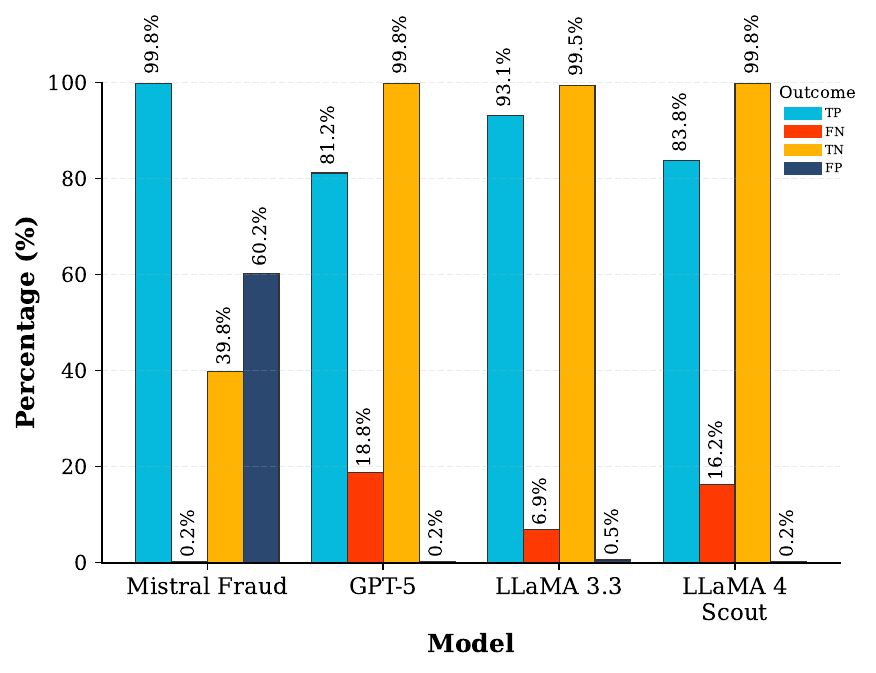}
\caption{LLMs exhibit mixed performance when classifying messages as `scam' or `not scam'. 
Mistral Fraud has the highest FP rate, incorrectly labeling 60.2\% of benign messages as scams, 
while GPT-5 has the highest FN rate, misclassifying 18.8\% of scam messages as benign. 
LLaMA~3.3 performs best overall, with a FN rate of 6.9\% and a FP rate of 0.5\%.}
\label{fig:llm_text_scale}
\end{figure}

Following an increasing body of prior work adopts LLMs for organizing and categorizing qualitative data~\cite{Liao_2024,Guo_2025}, we also evaluate the binary text classification performance of four off-the-shelf LLMs on mobile messaging scam texts and a benign baseline, reflecting how a typical end-user with limited technical skills would prompt LLMs for assistance. We focus on GPT-5, LLaMA 3.3, LLaMA 4 Scout, and Mistral Fraud models. We focus on GPT-5, LLaMA 3.3, LLaMA 4 Scout, and Mistral Fraud. We evaluate these models on our full scam dataset and Alexa’s Topical Chat dataset, which contains over 230,000 benign human text conversations~\cite{alexa_tc}.

\subsubsection{LLM Prompting and Datasets}
\label{subsec:prompting}
For our text detection experiments, we apply zero-shot prompting using text from our scam dataset. This approach is the most relevant evaluation strategy in this context reflecting realistic end-user behavior: non-expert users are likely to simply paste suspicious text into a model like ChatGPT and ask whether it is a scam without further finetuning~\cite{Zam_2023}. We evaluate both LLaMA 3.3 and LLaMA 4 Scout to compare accuracy across similar model families, motivated by prior work showing that LLaMA 4 Scout can underperform on text classification despite being newer and larger~\cite{Guo_2025}. Comparing these models further enables us to examine performance across widely varying parameter scales (e.g., LLaMA 3.3 uses 120 GB and runs on a single B200 GPU, while LLaMA 4 Scout requires over 200 GB and two GPUs). We select LLaMA 4 Scout rather than the full LLaMA 4 Maverick model because the latter requires significantly greater computational resources. We include Mistral Fraud because it is fine-tuned on synthetically generated fraudulent transcripts and is explicitly designed for fraud detection tasks~\cite{mistral_fraud}. GPT-5 is included to test the performance of popular, state-of-the-art models. Furthermore, unlike the other four models, which are hosted locally, GPT-5 is tested using Microsoft Azure's OpenAI API, which allows us to assess the strengths and weaknesses of commercial versus open-source LLM detection approaches. All models were given the following simple task prompt: \texttt{"Does this text look like a scam or not? Simply respond with 'SCAM' or 'NOT SCAM'."} 


\begin{figure}
    \centering
    \includegraphics[scale=0.56]{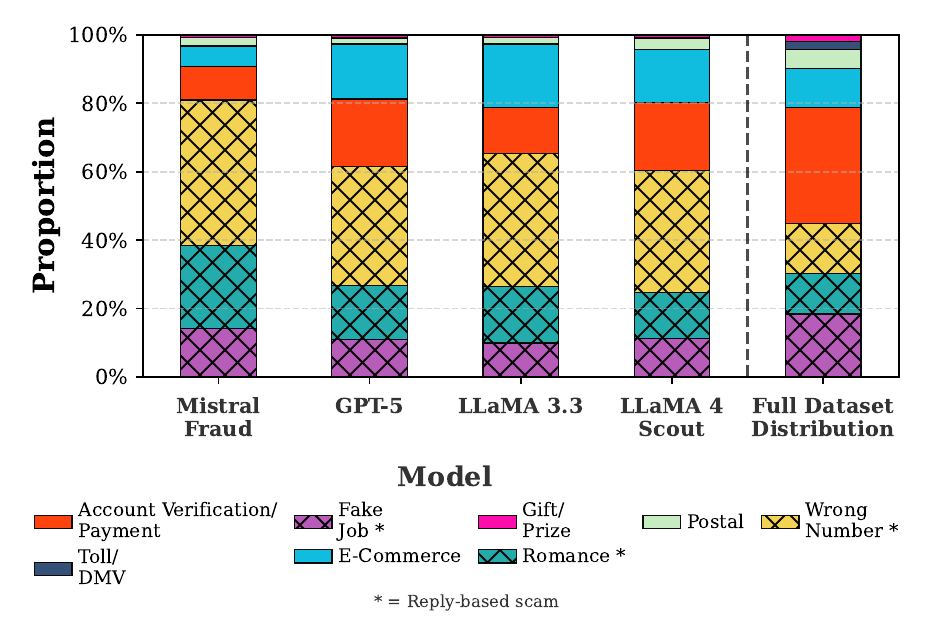}
    \caption{FN proportions for model in comparison with the full distribution of scams in the dataset on the right. Wrong Number scams are the most incorrectly classified category of scams across all models.}
    \label{fig:proportions}
\end{figure}

\subsubsection{LLM Results}
Using a simple prompt to emulate a typical end-user, defined in Section~\ref{subsec:prompting}, the model text classification output varies, as shown in Figure~\ref{fig:llm_text_scale}. Mistral Fraud (a model specifically fine-tuned to detect fraudulent messages) exhibits the worst precision with an extremely high FP rate of 60.2\%, making it unusable in practice. Conversely, both LLaMA models and GPT-5 show opposite results: high FN rates and relatively low FP rates, with LLaMA 4 Scout producing more than double the FNs of LLaMA 3.3. This suggests that LLaMA 3.3 is more capable at avoiding FNs than LLaMA 4 Scout, but both still miss many scams.

Although its results are similar to those of LLaMA 4 Scout, GPT-5 displays another distinct pattern. The API refuses to evaluate close to 4\% of scam messages (6,149 out of 175,430 total scam messages) due to ``inappropriate content’’ (e.g., profanity). This means that adding foul language is a trivial evasion strategy against commercial language models, as an adversary could insert profanity into the initial scam message to push the model into a non-decisive state.

To better understand areas of text misclassification, we analyze FNs (i.e., misclassified scam messages) by category. Figure~\ref{fig:proportions} reveals that LLMs do not uniformly misclassify scam messages by scam type. For GPT-5 and LLaMA models, several scam categories are misclassified, while the Mistral Fraud model misclassified the most reply-based scams (i.e., Wrong Number) and the fewest click-based scams (but suffers from high FPs). Wrong Number scams are the most misclassified scams across all models due to the diversity of topics and reliance on user replies, which causes messages to ultimately deviate from a few repeated scripts. As demonstrated in Section~\ref{sec:content_analysis}, click-based scams are, in many cases, easily detectable without the aid of language models capable of sophisticated reasoning due to their highly templated text content. Figure~\ref{fig:proportions} shows that several click-based scams, such as Postal and Toll/DMV scams, are among the most correctly classified categories in our dataset. This suggests that highly templated scams are already effectively handled by current LLMs and do not require additional attention. In contrast to this finding, LLMs across the board struggle with classifying reply-based Romance and Wrong Number scams, indicating that detection strategies relying on LLMs must be further refined before they can be considered reliable tools. Note also that Wrong Number scams constitute a smaller proportion of LLaMA 4's FN classifications as compared to LLaMA 3.3, but that LLaMA 4's overall FN rate is substantially higher than LLaMA 3.3's, indicating a potential tradeoff between ability to classify complex reply-based scams and overall detection performance.

Our findings indicate that while LLMs can accurately identify some mobile messaging scam content, no tested LLMs achieve the required performance for deployment. FP and FN error rates remain significantly higher than is typically tolerated in production spam filtering systems, where both rates are typically maintained $< $1\%~\cite{Delany_2012,Cormack_2007}. This gap highlights a limitation in current detectors: while LLMs can demonstrate strong recall in identifying scams, they remain overly sensitive, frequently misclassifying benign messages as malicious. Such behavior would be unacceptable in practical deployment contexts, where excessive FPs can degrade user trust and disrupt legitimate communications~\cite{Almeida_2011b}.

Overall, while LLMs show some promise for automated scam detection, their precision remains insufficient for real-world deployment in mobile messaging contexts. Performance varies substantially across scam categories: well-structured, link-centric scams (e.g., postal or toll notifications) are often detected reliably, whereas conversational, reply-driven scams remain challenging. Consequently, future improvements should prioritize these harder-to-detect categories rather than uniformly targeting all scam types, potentially through domain-specific fine-tuning or hybrid architectures that combine LLM reasoning with conventional filtering heuristics~\cite{Jiang_2024}. These findings are consistent with prior work showing that LLMs struggle to detect messaging scams under zero-shot prompting conditions~\cite{Schwarz_2025}.

\begin{tcolorbox}[standard jigsaw, opacityback=0,colframe=cyan]
    \textbf{RQ3.2 Takeaway:} \textit{LLMs do not perform consistently in detecting fraudulent messages, suffering from high FP rates and clear limitations in their ability to classify semantically-complex reply-based scams with and without context-specific fine-tuning.}
\end{tcolorbox}

%% file: limitations.tex
\section{Discussion and Limitations}
\label{sec:limitations}
In this section, we discuss the implications of our findings and outline key limitations that affect their interpretation.

\subsection{Discussion}
Although we observe clear trends and identifiable structure across the examined scams from Section~\ref{sec:background}, we find that the vast majority of scams ($>$95\% of our dataset) are similar at the level of semantic content but not sentence structure. This suggests that automated scam detection will continue to face challenges, especially as scammers apply new tools (e.g., LLMs) to customize content. We also observe cross-category structural variability: for example, the largest reply-based categories (i.e., Romance and Wrong Number) have no discernible high-structure subtypes.

We observe a sharp rise in reply-based scams between 2022 and 2023, coinciding with the widespread availability of consumer-facing LLMs such as ChatGPT~\cite{Marr_2023}. While prior work highlights the continued role of human labor in scam operations~\cite{Robles_2023}, such campaigns may plausibly combine human operators with LLM-assisted scripting and automation~\cite{Nield_2025}. Although we cannot distinguish human- from LLM-generated content or quantify LLM influence, our findings should be interpreted in a context where LLM-generated text is increasingly easy to produce yet difficult to reliably detect~\cite{Layton_2026}, limiting the effectiveness of text-based defenses.

This same period also coincides with the growing availability of phishing kits (collections of resources used to easily create and deploy phishing sites and messages at low costs~\cite{Oest_2018}) making it easier for less sophisticated attackers to create smaller campaigns~\cite{Resecurity_2023}. In 2025, it was discovered these same kits operate at a large scale, affecting at least 121 countries across numerous industries~\cite{Krebs_2025}.

Finally, we observe several numbers originating from non-U.S. countries, most notably the Philippines and Nigeria, particularly in the Postal and Romance categories. These findings align with recent evidence that suggests organized and centralized operations, such as a Philippines-based technology company providing internet domains (a core resource in click-based scams) to hundreds of thousands of scam platforms targeting victims globally~\cite{Kan_2025}. The aforementioned countries also predominantly speak English in neutral tones, further suggesting that English-speaking countries may see increasing attack volume.

\subsection{Limitations}
Our dataset is derived entirely from Reddit, discussed in Section~\ref{sec:methodology}, which introduces selection and platform-specific biases~\cite{brown2018reddit}. As a social media platform, Reddit reflects what users choose to share rather than the full spectrum of scam activity. Content visibility is driven by engagement, which biases the dataset toward scams that are novel, emotionally salient, or visually striking, while more routine or subtle scams are underrepresented~\cite{li2024like}. Although direct data collection (e.g., honeypots or provider-side measurements) could mitigate these biases, such approaches are constrained by legal restrictions on soliciting fraudulent communications and by widespread encryption in messaging platforms~\cite{requirements}. As a result, Reddit provides the most feasible and accessible data source for large-scale measurement.

Reddit’s user demographics also narrow the scope of scams included in our analysis. The platform primarily consists of younger English-speaking individuals from the United States and Canada (i.e., NANP countries), which may overlook scams targeting older populations, less active online individuals, and regions outside of North America~\cite{medvedev2017anatomy, houtti2024survey}. While we identify several non-NANP sources of scam activity (e.g., the Philippines, Nigeria, Thailand), our dataset primarily reflects English-language content within the NANP region. We also observe a small subset of non-English text conversations (e.g., Chinese, Spanish). We omit these texts from our analyses, as they represent only $< 1\%$ of our full dataset, and would not be representative of trends and international patterns. As such, we cannot yet capture predictive or linguistic patterns in non-English or non-U.S. messaging environments, limiting the generalizability of our findings. We also do not incorporate social media platforms outside the NANP ecosystem. Scam ecosystems in regions outside of the NANP operate in different linguistic, cultural, and regulatory contexts, and often involve scam delivery vectors, payment rails, and social markers that are not directly comparable to North American messaging norms (e.g., NANP telephone formats, payment infrastructure, slang in messages). As a result, extending this analysis to non-U.S. platforms would require collecting and analyzing entirely different social cues, narrative structures, and mobile messaging scam patterns~\cite{Shuter_2010}. This represents a distinct measurement study outside the scope of this work. 

Finally, several technical factors also limit our analyses. Reply-based scams often use encrypted messaging services (e.g., WhatsApp, Signal, Telegram) to converse with victims~\cite{Agarwal_2025b}, preventing visibility into off-platform interactions. Additionally, many of the phone numbers in our dataset are censored or redacted, limiting our ability to determine carrier or country origins more accurately. These constraints affect our capacity to fully characterize transnational messaging scams or quantify cross-border coordination in scam campaigns. As a limitation of Reddit, users often choose to redact and crop photos to preserve privacy, effectively removing messaging environment information and some phone numbers. Additionally, colors and fonts can be easily configured, based on the users' preferences, making it even more difficult to identify the image conversation's platform. Due to the prevalence of these occurrences, we could not reliably distinguish between different messaging environments. 

%% file: relatedwork.tex
\section{Related Work}
\label{sec:relatedwork}
Prior work on fraud detection and characterization has primarily focused on classifying scams using machine learning models~\cite{Salman_2022} and identifying content- or behavior-based signals within specific fraud schemes~\cite{Li_2018}. This literature spans a range of scam types, from general fraud detection to category-specific studies, most commonly phishing~\cite{Sheng_2009,Mishra_2020} and impersonation scams~\cite{Li_2018}.

Despite this body of work, automated scam detection remains relatively limited. Existing approaches often rely on Transformers~\cite{sbert} or hybrid neural models~\cite{Chen_2025,Schwarz_2025} and are applied primarily to phishing~\cite{Almeida_2011b,Gao_2012} and impersonation messages~\cite{Gupta_2018}. While effective for detection, these methods are typically constrained by dataset size and emphasize binary classification~\cite{Salman_2022}, limiting their ability to capture the operational diversity and structural behaviors of scam campaigns.

To supplement these approaches, research has also explored content-based detection, examining textual features such as spelling errors and embedded links~\cite{Afroz_2011}, as well as fraud signals in social networks and user profiles~\cite{Thomas_2011,Li_2024}. Research on mobile fraud and phishing similarly emphasizes anomalies in message content~\cite{Page_2018} and user interactions~\cite{Mishra_2020}, including analyses of URLs~\cite{Kambar_2023,Bitaab_2023}, lexical cues in impersonation scams~\cite{Agarwal_2025}, and message structure across scam and phishing categories~\cite{Swetha_2025,Agarwal_2025b,Khonji_2013}. Other studies incorporate behavioral signals such as call patterns and contact frequency~\cite{Li_2018}, or evaluate the effectiveness of blocklists~\cite{Sheng_2009,Peng_2019}. However, these efforts largely focus on detecting fraud rather than understanding how messaging scams are structurally organized, and few systematically compare scam categories using real-world operational characteristics.

Beyond detection, a smaller body of work examines the broader ecosystem and behavioral context of mobile fraud. Studies of spearphishing show that messages span diverse topics and impersonation strategies, such as posing as airline employees or insurance agents~\cite{Liu_2021}. Temporal analyses indicate that most phishing incidents occur on weekdays, particularly early in the week~\cite{Ho_2019}. Other work analyzes keyword usage, fraudulent usernames, and cryptocurrency-related patterns in YouTube comments~\cite{Li_2024,Zhang_2007}, or highlights the abuse of public SMS gateways in facilitating fraud~\cite{Reaves_2016,Nahapetyan_2024}. Additional studies characterize SMS scams through analyses of domain usage and lures~\cite{Agarwal_2025,Agarwal_2025b}, or focus on victim support by categorizing scams reported on online platforms~\cite{Sims_2025,Oak_2025a}. While informative, these studies either take a broad view that collapses multiple scam types into a single category~\cite{Reaves_2016,Nahapetyan_2024}, focus on a single platform~\cite{Agarwal_2025,Agarwal_2025b}, or examine narrowly defined scam categories~\cite{Liu_2021,Ho_2019}, limiting their generalizability across the broader fraud landscape.

Our work addresses these gaps by collecting a large corpus of real-world fraudulent mobile messages and systematically characterizing scams based on content, structural patterns, and message intent. This approach enables comparative analysis across scam types, revealing operational similarities and differences that are often obscured in detection-focused studies.

%% file: conclusion.tex
\section{Conclusion}
\label{sec:conclusion}
This paper presents a comprehensive real-world characterization of mobile messaging scams, integrating content analysis, attribute extraction, and evaluation of off-the-shelf defenses. Our results show that scam campaigns are diverse and dynamic: reply-based scams, such as Wrong Number and Romance, grow nearly twice as quickly as click-based scams and exhibit greater message diversity, reflecting a shift toward conversational engagement. We also find that specific campaigns are repeatedly associated with phone numbers from the same countries, suggesting persistent links between scam type and underlying infrastructure.

We further evaluate two detection layers, URL-based defenses and LLM-based classifiers, and find substantial limitations in both. URL defenses struggle with domain diversity and widespread use of shorteners, while LLMs fail to reliably classify scam texts, particularly with profanity or conversational cues. Together, these findings indicate that current detection tools are poorly aligned with the evolving structure of mobile messaging scams.

As scam tactics continue to evolve, our results underscore the need for more adaptive and interpretable detection frameworks that integrate content, behavioral, and attribute-level signals, rather than relying solely on static text or rule-based filtering.

%% file: appendix.tex
\section*{Appendix}
\input{open_science}
\section{Clustering and Scam Categories}
\label{appendix:clustering}
To encode our preprocessed lemmatized data into 384-dimensional vector representations, we use the all-MiniLM-L6-v2 model from the SentenceTransformer library. This model captures both syntactic and semantic features of short-text data, which is useful in analyzing the often fragmented and context-dependent nature of SMS messages. To improve clustering, we apply a multi-step dimensionality reduction pipeline: first, principal component analysis (PCA) is applied for linear dimensionality reduction, then we apply Uniform Manifold Approximation and Projection (UMAP)~\cite{umap} to nonlinearly reduce the high-dimensional embeddings into a smaller space while maintaining the semantic similarities between messages. We perform a grid search to determine the optimal k-means clustering parameters. We determine the optimal number of clusters using a combination of the elbow method and silhouette scores, shown in Appendix~\ref{appendix:tables}, Table~\ref{tab:cluster_counts}. To address small and/or fragmented clusters, we further refine clusters through manual merging of small clusters with their closest semantic neighbors based on centroid distance in the embedding space. In addition, 2 researchers manually merged similar clusters through independently labelling and then coming together to compare results. This resulted in $> 90\%$ agreeance for each of the 50 manually reviewed rows in each cluster, ending in the final clusters/categories used in this work. 


When subclustering, HDBSCAN is applied to sentence embeddings generated from cleaned, non-lemmatized text using the all-mpnet-base-v2 SentenceTransformer model, which generates 768-dimensional embeddings. Higher-dimensionality embeddings were used since the non-lemmatized data is likely to be of higher dimensionality. This methodology effectively accounts for intra-cluster stratifications, assisting in the detection of distinct subcategories within each cluster, which traditional means pairwise cosine similarity analysis would not capture. Additional keyword-based cluster merging was performed duringsub-clustering, revealing pockets of certain scams (e.g., postal and gift card scams) that would sometimes appear in the incorrect cluster, which were later merged to their correct cluster.


\section{Methodology}
\label{appendix:methodology}
\subsection{Phone Number and URL Matching}
We use the following regular expression to extract and normalize phone numbers found in our dataset:
\begin{quote}
\noindent\texttt{%
\textbackslash+?\textbackslash d[\textbackslash d\textbackslash s\textbackslash-()]%
\{7,\}\textbackslash d%
}
\end{quote}

\begin{figure}[!htbp]
    \centering
    \includegraphics[scale=0.53]{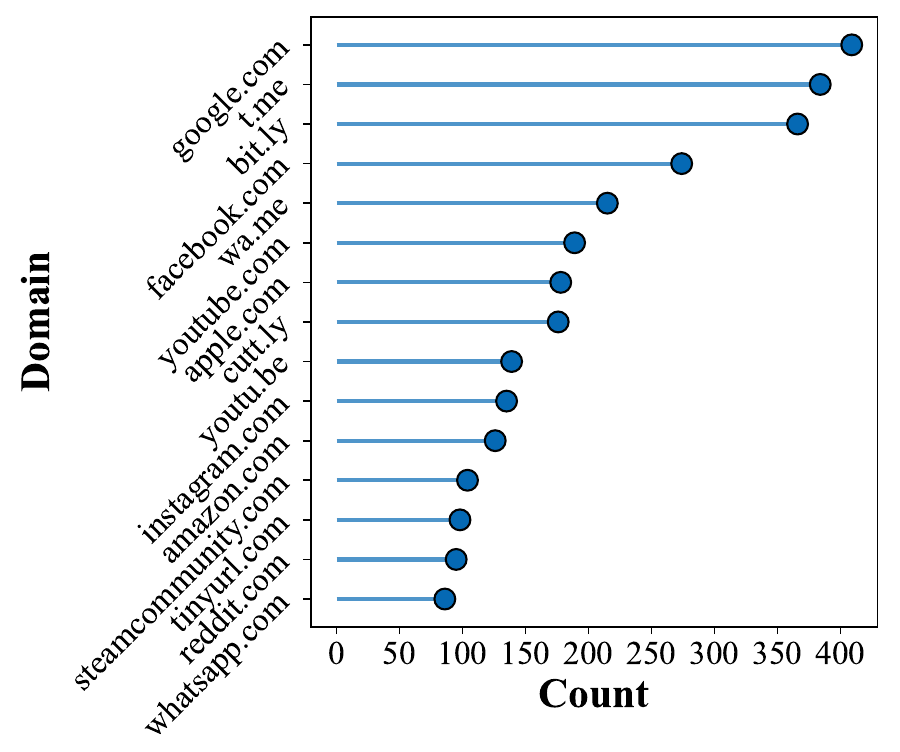}
    \caption{Domains are diverse and messages often use link shorteners.}
    \label{fig:domains}
    \vspace{-0.23in}
\end{figure}

This pattern matches strings of at least seven digits that do not include any non-number characters, excluding parentheses, dashes, and spaces, with an optional `+' at the beginning of the string. We perform additional processing using Python's phonenumbers library to ensure that extracted strings are legitimate phone numbers.
We use the following regular expression pattern to extract URLs: 

{\centering
\begin{quote}
\noindent\texttt{%
\textbackslash b((?:https?://|www\.)[a-zA-Z0-9.-]+\\.%
[a-zA-Z]\{2,6\}(?:/[\textasciicircum\textbackslash s,;:()<>"]*)?)|
\textbackslash b([a-zA-Z0-9.-]+\\.[a-zA-Z]\{2,6\}/[\textasciicircum\textbackslash s,;:()<>"]*)%
}
\end{quote}}
This pattern captures common URL/web link patterns, excluding email addresses, URLs containing spaces, and some incomplete links (though links beginning with only `www' and bare domains with paths are captured by the expression). Additionally, domains have several different URLs, such as longer and shorter versions of URLs (e.g., t.me and telegram.org are Telegram, youtu.be and youtube.com are YouTube); our regular expression captures these varying domain names. This process extracted 6,708 total phone numbers and 9,821 total URLs from our dataset, with the most common URLs shown in Figure~\ref{fig:domains}.

\subsection{Image Filtering}
We evaluated our filtering pipeline on 100 randomly sampled screenshots from the full unfiltered image set and correctly classified 95 images. Because screenshot layouts vary substantially (e.g., partial/cropped text message conversation screenshots, WhatsApp versus iMessage UI layouts), we adopted a conservative filtering strategy to retain only the most relevant images. Specifically, we reject images smaller than 300×300 pixels or with sharpness below 50, since low-quality images are not reliably processed by \texttt{pytesseract}. In addition, images with no text are also automatically discarded (e.g., selfies, pictures). To distinguish emails from text-message screenshots, we use a multi-step filter that checks for chat bubbles and chat-style UI cues, while separately detecting email indicators such as headers, email addresses, URLs, app keywords (e.g., inbox, drafts, archive, forward), and longer paragraph or bulleted-list structure in the message body.



\section{Domains}
\label{appendix:domains}
This section lists supplementary figures used in Section~\ref{sec:url_analysis}. These figures further describe the most prevalent links and scam categories associated with them.


\section{Tables}
\label{appendix:tables}

\begingroup
\renewcommand{\arraystretch}{0.9}    
\setlength{\tabcolsep}{4pt}           

\input{cluster_table}
\input{year_category_table}
\input{text_table}

\endgroup


%% file: open_science.tex
\section*{Open Science}
\label{sec:open_science}

\subsection*{Dataset}
To support reproducible science~\cite{Olszewski_2023}, we provide our dataset and all scripts available: \url{https://github.com/mobile-scam-analysis/mms-characterization}.

\subsection*{Ethical Considerations}
This study uses scam-related text messages collected from Reddit. While this data is publicly accessible and contains no personally identifiable information (PII), we acknowledge that user-generated content still requires ethical considerations~\cite{Bartlett_2025,Fiesler_2024}. We list the following steps taken to minimize harm and uphold ethical research practices, following guidelines provided specifically for using Reddit data in research~\cite{Fiesler_2024}:

\begin{itemize}
    \item \textbf{Context:} The Reddit posts used in this study were sourced from publicly accessible datasets from subreddits where users voluntarily shared scam messages that they received. These posts were submitted with the intention of warning others, to share experiences, or to discuss the nature and content of scams. While this data is public, we acknowledge that contributors may not expect their posts to be used in large-scale academic analyses. We have taken steps to respect this by avoiding intentionally collecting any user-level heuristics and focusing only on the textual content of the messages themselves and other features relevant to scams (i.e., phone numbers, URLs).
    \item \textbf{Consent and Data Minimization:} We did not interact with users. No usernames or profile metadata were collected or stored. Our dataset consists solely of the scam message texts as posted, with no attempt to identify characteristics about the individuals posting them.
    \item \textbf{Risk and Harm Mitigation:} To mitigate risks of reputational harm or re-identification of users, we did not reproduce full posts or usernames in the paper. All examples provided are paraphrased or anonymized, including messages used in Table~\ref{tab:reuse}.
\end{itemize}

%% file: cluster_table.tex
\begin{table}[H]
\centering
\begin{tabular}{l r}
\hline
\textbf{Cluster Name} & \textbf{Count} \\
\hline
Romance & 42,176 \\
Fake Job & 15,569  \\
Wrong Number & 19,397 \\ 
Account Verification/Payment &  42,148 \\
E-Commerce & 9,336 \\
Gift/Prize & 4,695 \\
Postal &  5,153 \\
Toll/DMV & 1,510 \\
\hline
\end{tabular}
\caption{Message count by cluster category.}
\label{tab:cluster_counts}
\end{table}

%% file: year_category_table.tex

\begin{table}[H]
\centering
\small
\begin{tabular}{p{2.5cm} rrrrrr}
\toprule
\textbf{Category} & \textbf{2020} & \textbf{2021} & \textbf{2022} & \textbf{2023} & \textbf{2024} & \textbf{2025} \\
\midrule
Fake Job                        &   615 &  1703 &  3207 &  7619 &  5776 &  9145 \\
Romance                         &   275 &   760 &  1325 &  2678 &  2050 & 39162 \\
Wrong Number                    &   475 &  1171 &  2124 &  5223 &  5586 &  9816 \\
Account Verification/Payment    &  2411 &  4464 &  8548 &  9087 &  7714 & 30177 \\
E-Commerce                      &   877 &  1502 &  2524 &  2648 &  3903 &  3506 \\
Gift/Prize                      &   136 &   236 &   340 &   458 &   416 &  4279 \\
Postal                          &     0 &     0 &     0 &  1830 &  3267 &  1267 \\
Toll/DMV                        &     0 &     0 &     0 &    25 &   278 &  1232 \\
\bottomrule
\end{tabular}
\caption{Yearly distribution of scam categories in the dataset.}
\label{tab:category_year}
\end{table}

%% file: text_table.tex
\definecolor{basecolor}{HTML}{2b4871}

\begin{table*}[]
\centering
\rowcolors{2}{basecolor!10}{white} 
\renewcommand{\arraystretch}{1.8} 

\begin{tabular}{l l p{8.5cm} r}
\toprule
\textbf{Category} & \textbf{Subcategory} & \textbf{Text} & \textbf{Count} \\
\midrule
Postal & -- -- & ``[USPS/Postal Service/Royal Mail/Canada Post Office/EVRi] package has arrived at the warehouse and cannot be delivered due to incomplete address information, Please confirm your address in the link. [link]" & 2,646 \\

Gift/Prize & Muse Scam & ``Hey, I really like your profile and your posts, and if you let me, I would make such an amazing mural out of it! If you don't mind one of your posts could be my inspiring muse for an art project I'm working on for a client. You will totally get paid for it as well as a bonus also get credits." & 679 \\

Toll/DMV & -- --  & ``Your vehicle has an unpaid toll bill. To avoid excessive late fees on your bill, please settle it promptly, Thank you for your cooperation! Total amount: [value]" & 123 \\

Gift/Prize & Reddit Modmail Scam & ``Hello, [username]! I found your new subreddit, I think you can achieve good results! I also had my own subreddit, but unfortunately it was banned. This happened because I created my own dating service (18+). Follow the link, you won't regret it. Here is the link: [link] (this is not a scam)'' & 108 \\

Account Payment/Verification & Bill Paid Scam & ``Free Msg: Your bill is paid for March. Thanks, here's a little gift for you: [link]" & 73 \\

Gift/Prize & Illuminati Scam & ``I am inviting you to join the Great Illuminati Organization now to become a billionaire, for fame, power, business, lucrative position, each new member will receive \$50,000,000.00 USD as benefit and \$50,000.00 USD as monthly payment and a new home in any country of their choice. and a model Mercedes Benz-SUV 2022, if you are interested, send your reply to: [email] NOTE: ALL MAIL SHOULD BE SENT TO: [email]." & 72 \\

Fake Job* & Part Time Job Scam & ``Welcome! We noticed that your background and resume have been recommended by several online recruitment agencies. That's why we want to offer you a part-time job that you can do in your free time. Our job is simple: we just review your favorite hotels. There is no time limit and you can complete the assessment at home. Daily wages range from \$300 to \$1,000, with all payments made on the same day. You can receive your salary immediately after each work day. If you would like to participate, please contact us via Whatsapp [number] (Note: you must be over 23 years old)" & 57 \\

Toll/DMV & FastTrak Lane Scam & ``Please pay for FastTrak Lane on December [number], 2024. In order to avoid excessive late fees and potential legal action on the bill, please pay the fee in time. Thank you for your cooperation and wish you a happy holiday. [link]'' & 58 \\
\end{tabular}

\caption{Most common scripts from scam campaigns across categories. USPS scams exhibit the highest uniformity, with similar repetition across other subcategories. Asterisk (*) denotes reply-based scams.}
\label{tab:reuse}
\end{table*}